\documentclass[aps,pra,reprint,groupedaddress]{revtex4-1}
\usepackage{amssymb}
\usepackage{amsmath}
\usepackage{array}
\usepackage{multirow,bigdelim}
\usepackage{enumitem}
\usepackage{graphicx}
\usepackage{epsfig}
\usepackage{graphics}
\usepackage{latexsym}

\begin{document}

\title{ Detection of genuine n-qubit entanglement via the proportionality of
two vectors}
\author{ Dafa Li$^{1,2}$}

\begin{abstract}
In [Science 340, 1205, 7 June (2013)], via polytopes Michael Walter et al.
proposed a sufficient condition detecting the genuinely entangled pure
states. In this paper, we indicate that generally, the coefficient vector of
a pure product state of $n$ qubits cannot be decomposed into a tensor
product of two vectors, and show that a pure state of $n$ qubits is a
product state if and only if there exists a permutation of qubits such that
under the permutation, its coefficient vector arranged in ascending
lexicographical order can be decomposed into a tensor product of two
vectors. The contrapositive of this result reads that a pure state of $n$
qubits is genuinely entangled if and only if its coefficient vector cannot
be decomposed into a tensor product of two vectors under any permutation of
qubits. Further, by dividing a coefficient vector into $2^{i}$ equal-size
block vectors, we show that the coefficient vector can be decomposed into a
tensor product of two vectors if and only if any two non-zero block vectors
of the coefficient vector are proportional. In terms of \textquotedblleft
proportionality\textquotedblright , we can rephrase that a pure state of $n$
qubits is genuinely entangled if and only if there are two non-zero block
vectors of the coefficient vector which are not proportional under any
permutation of qubits. Thus, we avoid decomposing a coefficient vector into
a tensor product of two vectors to detect the genuine entanglement. We also
present the full decomposition theorem for product states of n qubits.
\end{abstract}


\affiliation{
$^1$Department of Mathematical Sciences, Tsinghua University,
Beijing, 100084, China\\
$^2$Center for Quantum Information Science and Technology, Tsinghua National
Laboratory for Information Science and Technology (TNList), Beijing,
100084, China}


\maketitle

\section{Introduction}

Quantum entanglement is a crucial physical resource in quantum information
theory: quantum communication, quantum computation, quantum cryptography,
quantum teleportation, quantum dense coding, and so on \cite{Nielsen}. A
quantum state of $n$ qubits\ is genuinely entangled\ if it cannot be written
as $|\varphi \rangle |\phi \rangle $. Detecting if a state is genuinely
entangled\ is vital in quantum information theory and experiments \cite%
{Horodecki, Li-Pan}. Whereas, detecting and measuring genuine entanglement
has turned out to be a challenging task.

In \cite{Peres}, Peres has presented a positive partial transpose (PPT)
criterion\ for\ a separability of bipartite systems. To detect multi genuine
entanglement, previous articles proposed linear and nonlinear entanglement
witnesseses \cite{Huber10}-\cite{Jungnitsch}, Bell-like inequalities \cite%
{Bancal}, and generalized concurrence \cite{Ming-Li-15, Chen, Gao}. In \cite%
{Fei-sci-17, Ming-Li, Ming-Jing, Ming-Jing-11, Shen, DLi-CTP}, sufficient
conditions to detect and measure genuine bipartite and tripartite\
entanglement for pure states were presented. Very recently, a general
criterion was proposed for detecting multilevel entanglement in
multiparticle quantum states \cite{Kraft} and a necessary and sufficient
criterion for the separability of bipartite mixed states \cite{Jun-Li}. In
\cite{Walter},\ via Polytopes and linear inequalities describing eigenvalues
of single-particle density matrices, Michael Walter et al. gave the
sufficient condition for the genuinely entangled pure states. Specially, up
to permutation of the particles, they showed that 7 of 13 entanglement
polytopes belong to distinct types of genuine four-partite entanglement.

The previous articles proposed the conditions, which are not only necessary
but also sufficient, for the separability of pure states for bipartite \cite%
{Verstraete01}, multipartite \cite{Vicente}, and multipartite \cite{Bastin18}%
, and a necessary and sufficient condition for the full separability of pure
states of three qubits \cite{Yu}.

In this paper, we indicate that the coefficient vector of a pure product
state of $n$ qubits cannot be always decomposed, and we propose a necessary
and sufficient condition for detecting genuine entanglement of $n$ qubits\
via the tensor decomposition of the coefficient vectors under permutations
of qubits. In this paper, we also show that a coefficient vector can be
decomposed if and only if any two non-zero block vectors of a block matrix
of the coefficient vector are proportional. Thus, we give another necessary
and sufficient condition for the genuine entanglement via \textquotedblleft
proportionality\textquotedblright\ of two vectors\ and permutations of
qubits. We also demonstrate how to obtain a decomposition of an coefficient
vector of a pure state via \textquotedblleft
proportionality\textquotedblright\ of two vectors.. After then, we can
directly factorize the state. It means that we give a constructive method
for the decomposition of pure product states. Moreover, we derive the full
decomposition theorem.

\section{Detection of genuinely entangled pure states of $n$ qubits via the
tensor decomposition of the coefficient vectors}

Any pure state of n qubits can be written as $|\psi \rangle _{1\cdots
n}=\sum_{i=0}^{2^{n}-1}c_{i}|i\rangle $. Let the $2^{n}$ dimensional vector $%
C(|\psi \rangle _{1\cdots n})=(c_{0},c_{1},\cdots ,c_{2^{n}-1})^{T}$ whose
entries are the coefficients $c_{0},c_{1},\cdots ,c_{2^{n}-1}$ of the state $%
|\psi \rangle _{1\cdots n}$ arranged in ascending lexicographical order,
where $C^{T}$ is the transpose of $C$.\ We call $C(|\psi \rangle _{1\cdots
n})$ the coefficient vector of the state $|\psi \rangle _{1\cdots n}$.\ We
write $C(|\psi \rangle _{1\cdots n})$\ as $C(|\psi \rangle )$ sometimes.\ We
say that the coefficient vector $C(|\psi \rangle )$ can be decomposed if
\begin{equation}
C(|\psi \rangle )=V_{1}\otimes V_{2},
\end{equation}%
where $V_{1}$ is a $2^{i}$ dimensional vector and $V_{2}$ is a $2^{n-i}$
dimensional vector with $1\leq i\leq n-1$.

We define that a pure state $|\psi \rangle _{1\cdots n}$\ of $n$ qubits is a
product state if and only if $|\psi \rangle _{1\cdots n}=|\varphi \rangle
|\phi \rangle $.

\subsection{For two qubits}

For two qubits, any pure state can be written as $\sum_{i=0}^{3}c_{i}|i%
\rangle $. It is well known that a pure state of two qubits is a product
state if and only if $c_{0}c_{3}=c_{1}c_{2}$ \cite{LDFPRA}. One can check
that $c_{0}c_{3}=c_{1}c_{2}$ is equivalent to that the coefficient vector $%
\left(
\begin{array}{cccc}
c_{0} & c_{1} & c_{2} & c_{3}%
\end{array}%
\right) ^{T}$ can be decomposed. Next, we explore the decomposition of the
coefficient vectors.

If $|\psi \rangle _{12}$ is product, then we can write $|\psi \rangle
_{12}=|\varphi \rangle _{1}|\phi \rangle _{2}$, where $|\varphi \rangle
_{1}=\alpha |0\rangle _{1}+\beta |1\rangle _{1}$ and $|\phi \rangle
_{2}=\gamma |0\rangle _{2}+\delta |1\rangle _{2}$. A calculation yields that
$C(|\psi \rangle )=\left(
\begin{array}{cccc}
\alpha \gamma & \alpha \delta & \beta \gamma & \beta \delta%
\end{array}%
\right) ^{T}=C(|\varphi \rangle )\otimes C(|\phi \rangle )$, where $%
C(|\varphi \rangle )=(\alpha ,\beta )^{T}$ and $C(|\phi \rangle )=\left(
\begin{array}{cc}
\gamma & \delta%
\end{array}%
\right) ^{T}$.

Therefore, we can conclude that a pure state of two qubits is a product
state if and only if its coefficient vector can be decomposed.

\subsection{For three qubits}

Let $\rho _{123}$ be the density matrix of the state\ $|\psi \rangle _{123}$%
, i.e. $\rho _{123}=|\psi \rangle _{123}\langle \psi |$. A state of three
qubits is a product state if and only if the reduced densities $\rho _{1}$, $%
\rho _{2}$, or $\rho _{3}$ have rank 1, where $\rho _{1}=tr_{23}\rho _{123}$
\cite{Dur}. Next, we explore the decomposition of the coefficient vectors.

Assume that $|\psi \rangle _{123}$ is product. Then, there are three cases.

Case 1. $|\psi \rangle _{123}=|\varphi \rangle _{1}|\phi \rangle _{23}$,
where $\ |\varphi \rangle _{1}=(\alpha |0\rangle _{1}+\beta |1\rangle _{1})$
and $|\phi \rangle _{23}=(a|00\rangle _{23}+b|01\rangle _{23}+c|10\rangle
_{23}+d|11\rangle _{23})$.

For the case, a calculation yields that $C(|\psi \rangle )=C(|\varphi
\rangle )\otimes C(|\phi \rangle )$, where $C(|\varphi \rangle )=\left(
\begin{array}{cc}
\alpha & \beta%
\end{array}%
\right) ^{T}$ and $C(|\phi \rangle )=\left(
\begin{array}{cccc}
a & b & c & d%
\end{array}%
\right) ^{T}$.

Case 2. $|\psi \rangle _{123}=|\phi \rangle _{12}|\varphi \rangle _{3}$,
where $|\varphi \rangle _{3}=(\alpha |0\rangle _{3}+\beta |1\rangle _{3})$
and $|\phi \rangle _{12}=$ $(a|00\rangle _{12}+b|01\rangle _{12}+c|10\rangle
_{12}+d|11\rangle _{12})$.

For the case, a calculation yields that $C(|\psi \rangle )=C(|\phi \rangle
)\otimes C(|\varphi \rangle )$.

Case 3. $|\psi \rangle _{123}=|\varphi \rangle _{2}|\phi \rangle _{13}$,
where $|\varphi \rangle _{2}=\alpha |0\rangle _{2}+\beta |1\rangle _{2}$ and
$|\phi \rangle _{13}=a|00\rangle _{13}+b|01\rangle _{13}+c|10\rangle
_{13}+d|11\rangle _{13}$.

When $bc\neq ad$, i.e. $|\phi \rangle _{13}$ is a genuinely entangled
state,\ we can show that $C(|\psi \rangle )$ cannot be decomposed into $%
V_{1}\otimes V_{2}$ though $|\psi \rangle $ is a product state. Now, let us
consider the transposition $\pi =(1,2)$ of qubits 1 and 2. Then under the
transposition $\pi $, $|\psi \rangle _{123}$ becomes $|\psi ^{\prime
}\rangle _{123}=|\varphi ^{\prime }\rangle _{1}|\phi ^{\prime }\rangle _{23}$%
, where $|\varphi ^{\prime }\rangle _{1}=\alpha |0\rangle _{1}+\beta
|1\rangle _{1}$, and $|\phi ^{\prime }\rangle _{23}$ $=a|00\rangle
_{23}+b|01\rangle _{23}+c|10\rangle _{23}+d|11\rangle _{23}$. One can see
that $C(\psi ^{\prime })$ can be decomposed into $C(\psi ^{\prime
})=C(\varphi ^{\prime })\otimes C(\phi ^{\prime })$.

So, we can conclude that for three qubits, if $|\psi \rangle $ is a product
state then the coefficient vectors $C(|\psi \rangle )$ or $C(|\pi \psi
\rangle )$, where $\pi $ is the transposition $(1,2)$ of qubits 1 and 2, can
be decomposed into tensor products of two vectors.

We can generalize the result below.

\subsection{For $n$ qubits}

\textit{Lemma 1.} If a pure state $|\psi \rangle _{1\cdots n}$\ of $n$
qubits is a product state of a state $|\varphi \rangle _{1\cdots i}$ of
qubits $1,\cdots ,i$ and a state $|\phi \rangle _{(i+1)\cdots n}$ of qubits $%
(i+1),\cdots ,n$, i.e.

\begin{equation}
|\psi \rangle _{1\cdots n}=|\varphi \rangle _{1\cdots i}|\phi \rangle
_{(i+1)\cdots n},  \label{prod-0}
\end{equation}%
then the coefficient vector $C(|\psi \rangle )$\ can be decomposed into
\begin{equation}
C(|\psi \rangle )=C(|\varphi \rangle )\otimes C(|\phi \rangle ).
\label{g-n-1}
\end{equation}

Proof. Let $|\varphi \rangle _{1\cdots i}=\sum_{j=0}^{2^{i}-1}a_{j}|j\rangle
_{1\cdots i}$ and $|\phi \rangle _{(i+1)\cdots
n}=\sum_{k=0}^{2^{n-i}-1}b_{k}|k\rangle _{(i+1)\cdots n}$. Then, from Eq. (%
\ref{prod-0}) a calculation yields
\begin{eqnarray}
C(|\psi \rangle ) &=&(%
\begin{array}{cccc}
a_{0}b_{0} & a_{0}b_{1} & \cdots & a_{0}b_{2^{n-i}-1}%
\end{array}
\notag \\
&&%
\begin{array}{cccc}
a_{1}b_{0} & a_{1}b_{1} & \cdots & a_{1}b_{2^{n-i}-1}%
\end{array}
\notag \\
&&\cdots  \notag \\
&&%
\begin{array}{cccc}
a_{2^{i}-1}b_{0} & a_{2^{i}-1}b_{1} & \cdots & a_{2^{i}-1}b_{2^{n-i}-1}%
\end{array}%
)^{T}.  \label{vec-1}
\end{eqnarray}%
Eq. (\ref{vec-1}) implies that Eq. (\ref{g-n-1}) holds.

\textit{Lemma 2.} Let $|\psi \rangle _{1\cdots
n}=\sum_{i=0}^{2^{n}-1}c_{i}|i\rangle $ be any pure state of $n$\ qubits. If
the coefficient vector $C(|\psi \rangle )$\ can be decomposed into
\begin{equation}
C(|\psi \rangle )=V_{1}\otimes V_{2},  \label{cond-1}
\end{equation}%
where $V_{1}$ is a $2^{i}$ dimensional vector and $V_{2}$ is a $2^{n-i}$
dimensional vector with $1\leq i\leq n-1$, then we can define states $%
|\varphi \rangle _{1\cdots i}$ of qubits $1,\cdots ,i$, where $C(|\varphi
\rangle )=V_{1}$, and $|\phi \rangle _{(i+1)\cdots n}$ of qubits $%
(i+1),\cdots ,n$, where $C(|\phi \rangle )=V_{2}$, such that

\begin{equation}
|\psi \rangle _{1\cdots n}=|\varphi \rangle _{1\cdots i}|\phi \rangle
_{(i+1)\cdots n}.  \label{sep-1}
\end{equation}%
That is, the state $|\psi \rangle _{1\cdots n}$ is a product state.

Proof. Assume that $V_{1}=(a_{0},a_{1},\cdots ,a_{2^{i}-1})^{T}$ and $%
V_{2}=(b_{0},b_{1},\cdots ,b_{2^{n-i}-1})^{T}$. We define states $|\varphi
\rangle _{1\cdots i}=\sum_{j=0}^{2^{i}-1}a_{j}|j\rangle $ of qubits $%
1,\cdots ,i$ and $|\phi \rangle _{(i+1)\cdots
n}=\sum_{k=0}^{2^{n-i}-1}b_{k}|k\rangle $ of qubits $(i+1),\cdots ,n$.
Clearly, $C(|\varphi \rangle )=V_{1}$ and $C(|\phi \rangle )=V_{2}$. Via Eq.
(\ref{cond-1})\ a calculation yields Eq. (\ref{sep-1}). It means that $|\psi
\rangle _{1\cdots n}$ is a product state.

Lemmas 1 and 2 lead to the following Theorem 1.

\textit{Theorem 1}. A pure product state $|\psi \rangle _{1\cdots n}$\ of $n$
qubits is of the form%
\begin{equation}
|\psi \rangle _{1\cdots n}=|\varphi \rangle _{1\cdots i}|\phi \rangle
_{(i+1)\cdots n},  \label{separ-1}
\end{equation}%
where $i=1,2,\cdots ,n-1$, if and only if
\begin{equation}
C(|\psi \rangle )=V_{1}\otimes V_{2},  \label{tensor-eq-1}
\end{equation}%
where $V_{1}$ and $V_{2}$ are $2^{i}$ and $2^{n-i}$\ dimensional vectors,
respectively.

\textit{Lemma 3} (decomposition for product states). Let $|\varphi \rangle
_{k_{1}k_{2}\cdots k_{\ell }}$ be a state of qubits $k_{1},k_{2},\cdots
,k_{\ell }$, and $|\phi \rangle _{k_{(\ell +1)}\cdots k_{n}}$ be a state of
qubits $k_{(\ell +1)},\cdots ,k_{n}$. If \
\begin{equation}
|\psi \rangle _{1\cdots n}=|\varphi \rangle _{k_{1}k_{2}\cdots k_{\ell
}}|\phi \rangle _{k_{(\ell +1)}\cdots k_{n}},
\end{equation}%
then there exists a permutation of qubits, for example $\pi $, such that the
coefficient vector $C(|\pi \psi \rangle )$ of the state $|\pi \psi \rangle $%
\ is decomposed into
\begin{equation}
C(|\pi \psi \rangle )=C(|\pi \varphi \rangle )\otimes C(|\pi \phi \rangle ).
\label{g-n-3}
\end{equation}

Proof. Without loss of generality, assume that $\{k_{1},k_{2},\cdots
,k_{\ell }\}\cap \{1,\cdots ,\ell \}$ is the empty. Let the permutation $\pi
=(k_{\ell },\ell )\cdots (k_{2},2)(k_{1},1)$. Under the permutation $\pi $\
of qubits, $\{k_{1},k_{2},\cdots ,k_{\ell }\}$ and $\{k_{(\ell +1)},\cdots
,k_{n}\}$ become $\{1,\cdots ,\ell \}$ and $\{(\ell +1),\cdots ,n\}$,
respectively, and the states $|\psi \rangle _{1\cdots n}$, $|\varphi \rangle
_{k_{1}k_{2}\cdots k_{\ell }}$, and $|\phi \rangle _{k_{(\ell +1)}\cdots
k_{n}}$ become states $|\pi \psi \rangle _{1\cdots n}$, $|\pi \varphi
\rangle _{1\cdots \ell }$, and $|\pi \phi \rangle _{(\ell +1)\cdots n}$.
Thus,%
\begin{equation}
|\pi \psi \rangle _{1\cdots n}=|\pi \varphi \rangle _{1\cdots \ell }|\pi
\phi \rangle _{(\ell +1)\cdots n}.  \label{prod-1}
\end{equation}%
In light of Lemma 1, from Eq. (\ref{prod-1})\ we obtain Eq. (\ref{g-n-3}).

To understand Lemma 3, readers can refer Case 3 of the above subsection.

Lemmas 1, 2, and 3 lead to the following Theorem 2.

\textit{Theorem 2. }$|\psi \rangle _{1\cdots n}=|\varphi \rangle |\phi
\rangle $ if and only if there exists a permutation of qubits, for example $%
\pi $, such that the coefficient vector $C(|\pi \psi \rangle )$ can be
decomposed into
\begin{equation}
C(|\pi \psi \rangle )=V_{1}\otimes V_{2}.  \label{decom-1}
\end{equation}

Proof. Lemma 3 shows that if $|\psi \rangle _{1\cdots n}=|\varphi \rangle
\phi \rangle $, then there exists a permutation of qubits, for example $\pi $%
, such that $C(|\pi \psi \rangle )=C(|\pi \varphi \rangle )\otimes C(|\pi
\phi \rangle )$. Conversely, if there exists a permutation of qubits, for
example $\pi $, such that Eq. (\ref{decom-1}) holds. Assume that $V_{1}$ is
a $2^{i}$ dimensional vector and $V_{2}$ is a $2^{n-i}$ dimensional vector
with $1\leq i\leq n-1$. Then, in light of Lemma 2, we can define a state $%
|\varphi \rangle _{1\cdots i}$ of qubits $1,\cdots ,i$, where $C(|\varphi
\rangle _{1\cdots i})=V_{1}$, and a state $|\phi \rangle _{(i+1)\cdots n}$
of qubits $(i+1),\cdots ,n$, where $C(|\phi \rangle _{(i+1)\cdots n})=V_{2}$%
, such that
\begin{equation}
\pi |\psi \rangle _{1\cdots n}=|\varphi \rangle _{1\cdots i}|\phi \rangle
_{(i+1)\cdots n}.  \label{decom-3}
\end{equation}%
Let $\pi ^{-1}$ be the inverse of $\pi $. Under the permutation $\pi ^{-1}$,
the states $|\varphi \rangle _{1\cdots i}$ and $|\phi \rangle _{(i+1)\cdots
n}$ become states $|\pi ^{-1}\varphi \rangle _{\pi ^{-1}(1\cdots i)}$ and $%
|\pi ^{-1}\phi \rangle _{\pi ^{-1}((i+1)\cdots n)}$, respectively. Then,
from Eq. (\ref{decom-3}) we obtain
\begin{equation}
|\psi \rangle _{1\cdots n}=|\pi ^{-1}\varphi \rangle _{\pi ^{-1}(1\cdots
i)}|\pi ^{-1}\phi \rangle _{\pi ^{-1}((i+1)\cdots n)}.
\end{equation}%
\ \ Theorem 2 is an extension of Theorem 1.

\textit{Corollary 1. }Restated in the contrapositive Theorem 2 reads: A pure
state $|\psi \rangle _{1\cdots n}$\ of $n$ qubits is genuinely entangled if
and only if the coefficient vector $C(|\pi \psi \rangle )$ cannot be
decomposed into a tensor product of two vectors under any permutation $\pi $
in Appendix A.

For example, in light of Corollary 1 we can detect the following entangled
pure state of n qubits. For the GHZ\ state and W state of n qubits, $C(\pi $%
GHZ$)=C($GHZ$)$ and $C(\pi $W$)=C($W$)$ under any permutation $\pi $ of
qubits, and it is easy to check that $C($GHZ$)$ and $C($W$)$ cannot be
decomposed. Therefore, in light of Corollary 1, $n$-qubit GHZ and W are
genuinely entangled. Similarly, we can also test that Dicke states $%
|i,n\rangle $, GHZ+W, GHZ+Dicke states, and the state $\sum_{i_{1},\cdots
,i_{n}=0,1}|i_{1}i_{2}\cdots i_{n}\rangle -$ $|0\cdots 0\rangle -|1\cdots
1\rangle $ are genuinely entangled.

Let $|$DW$\rangle =$ $|1,n\rangle +|(n-1),n\rangle $.\ Next we demonstrate
that $|$DW$\rangle $ is genuinely entangled and does not belong to W SLOCC
(stochastic local operations and classical communication)\ class though $%
|1,n\rangle $ is the W state and $|(n-1),n\rangle $ belongs to W SLOCC\
class. The following is our argument. One can check that its coefficient
vector does not change under any permutation of qubits and the coefficient
vector cannot be decomposed. Therefore, in light of Corollary 1 this state
is genuinely entangled. The concurrence for $|$DW$\rangle $ of even $n$
qubits does not vanish while the concurrence vanishes for the W state of
even $n$ qubits \cite{LDFPRA}, so $|$DW$\rangle $ does not belong to W SLOCC
class for even $n$ qubits. The odd $n$-tangle does not vanish for $|$DW$%
\rangle $ of odd $n$ qubits while the odd $n$-tangle vanishes for the W
state of odd $n$ qubits \cite{LDFQIP}, so $|$DW$\rangle $ does not belong to
W SLOCC class for odd $n$ qubits. One can know that for three qubits, $|$DW$%
\rangle $ belongs to GHZ SLOCC\ class while for four qubits, a calculation
shows that it belongs to the family $L_{abc_{2}}$ \cite{Verstraete}.

Similarly, we can show that $|i,n\rangle +|(n-i),n\rangle $, where $%
i=2,\cdots ,[n/2]$, and $|i,n\rangle $ and $|(n-i),n\rangle $ are Dicke
states, are genuinely entangled.

\section{The full decomposition theorem}

In this section, we derive a full decomposition theorem. To that end, $n$
qubits are split into $m$ parts $R_{1}$, $R_{2}$, $\cdots $, $R_{m}$, where
\begin{eqnarray}
R_{1} &=&\{1,2,\cdots ,l_{1}\},  \notag \\
R_{2} &=&\{l_{1}+1,l_{1}+2,\cdots ,l_{1}+l_{2}\},  \notag \\
&&\cdots ,  \notag \\
R_{m} &=&\{n-(l_{m}-1),\cdots ,n-1,n)\}.  \label{set-1}
\end{eqnarray}

For example, six qubits are split into $R_{1}=\{1,2\}$, $R_{2}=\{3\}$, and $%
R_{4}=\{4,5,6\}$.

We generalize Lemma 1 below.

\textit{Lemma 4}. If a product state $|\psi \rangle _{1\cdots n}$ of $n$
qubits is of the form

\begin{equation}
|\psi \rangle _{1\cdots n}=|\psi _{1}\rangle _{R_{1}}|\psi _{2}\rangle
_{R_{2}}\cdots |\psi _{m}\rangle _{R_{m}},  \label{ten-1}
\end{equation}%
where $n$ qubits are split into $m$ parts $R_{1}$, $R_{2}$, $\cdots $, $%
R_{m} $ in Eq. (\ref{set-1}), then the coefficient vector $C(|\psi \rangle
_{1\cdots n})$ can be decomposed into
\begin{eqnarray}
&&C(|\psi \rangle _{1\cdots n})  \notag \\
&=&C(|\psi _{1}\rangle _{R_{1}})\otimes C(|\psi _{2}\rangle _{R_{2}})\otimes
\cdots \otimes C(|\psi _{m}\rangle _{R_{m}}).  \notag \\
&&  \label{g-n-5}
\end{eqnarray}

Proof. From Eq. (\ref{ten-1}), we can write $|\psi \rangle _{1\cdots
n}=|\varphi \rangle |\phi \rangle $, where $|\varphi \rangle =|\psi
_{1}\rangle _{R_{1}}|\psi _{2}\rangle _{R_{2}}\cdots |\psi _{m-1}\rangle
_{R_{m-1}}$ and $|\phi \rangle =|\psi _{m}\rangle _{R_{m}}$. In light of
Lemma 1, $C(|\psi \rangle )=C(|\varphi \rangle )\otimes C(|\psi _{m}\rangle
) $. Then, we can apply Lemma 1 to the state $|\varphi \rangle $. Then, by
the induction, we can obtain Eq. (\ref{g-n-5}).

\textit{Theorem 3} (The Full Decomposition Theorem) If a product state $%
|\psi \rangle _{1\cdots n}$\ \ of $n$ qubits\ is of the form
\begin{equation}
|\psi \rangle _{1\cdots n}=|\psi _{1}\rangle _{S_{1}}|\psi _{2}\rangle
_{S_{2}}\cdots |\psi _{m}\rangle _{S_{m}},  \label{decom-2}
\end{equation}%
where $n$ qubits are split into $m$ parts $S_{1}$, $S_{2}$, $\cdots $, $%
S_{m} $, then there exists a permutation of qubits, for example $\pi $, such
that the coefficient vector$\ C(|\pi \psi \rangle _{1\cdots n})$\ can be
decomposed into%
\begin{eqnarray}
&&C(|\pi \psi \rangle _{1\cdots n})  \notag \\
&=&C(|\pi \psi _{\ell _{1}}\rangle _{R_{1}})\otimes C(|\pi \psi _{\ell
_{2}}\rangle _{R_{2}})\otimes \cdots \otimes C(|\pi \psi _{\ell _{m}}\rangle
_{R_{m}}),  \notag \\
&&  \label{g-n-4}
\end{eqnarray}%
where $\pi S_{\ell _{k}}=R_{k}$, $k=1,\cdots ,m$, in Eq. (\ref{set-1}).

Proof. There exists a permutation of qubits, for example $\pi $, such that $%
\pi S_{\ell _{k}}=R_{k}$, $k=1,\cdots ,m$, in Eq. (\ref{set-1}), the states $%
|\psi \rangle _{1\cdots n}$ and $|\psi _{\ell _{k}}\rangle _{S_{\ell _{k}}}$
become states $|\pi \psi \rangle _{1\cdots n}$ and $|\pi \psi _{\ell
_{k}}\rangle _{R_{k}}$, respectively. Thus, under the permutation $\pi $,
Eq. (\ref{decom-2}) becomes
\begin{equation}
|\pi \psi \rangle _{1\cdots n}=|\pi \psi _{\ell _{1}}\rangle _{R_{1}}|\pi
\psi _{\ell _{2}}\rangle _{R_{2}}\cdots |\pi \psi _{\ell _{m}}\rangle
_{R_{m}}.
\end{equation}%
Then, in light of Lemma 4, Eq. (\ref{g-n-4}) holds.

\textit{Lemma 5. }For any pure state $|\psi \rangle _{1\cdots n}$\ of $n\ $%
qubits, if under a permutation of n qubits, for example $\pi $, the
coefficient vector$\ C(|\pi \psi \rangle _{1\cdots n})$ can be decomposed
into
\begin{equation}
C(|\pi \psi \rangle _{1\cdots n})=V_{1}\otimes V_{2}\otimes \cdots \otimes
V_{m},  \label{g-n-6}
\end{equation}%
then $|\psi \rangle _{1\cdots n}$ can be written as
\begin{equation}
|\psi \rangle _{1\cdots n}=|\psi _{1}\rangle _{S_{1}}|\psi _{2}\rangle
_{S_{2}}\cdots |\psi _{m}\rangle _{S_{m}},  \label{g-n-7}
\end{equation}%
where $n$ qubits are split into $m$ parts $S_{1}$, $S_{2}$, $\cdots $, $%
S_{m} $, and $|\psi _{i}\rangle _{S_{i}}$ is a state of qubits in $S_{i}$, $%
i=1,\cdots ,m$.

Proof. Assume that Eq. (\ref{g-n-6}) holds. Via the dimensions of $V_{i}$, $%
i=1,\cdots ,m$, we can split $n$ qubits into $m$ parts $R_{1}$, $R_{2}$, $%
\cdots $, $R_{m}$ in Eq. (\ref{set-1}). Then, we define a state $|\psi
_{i}^{\prime }\rangle _{R_{i}}$ of qubits in $R_{i}$ such that $C(|\psi
_{i}^{\prime }\rangle _{R_{i}})=V_{i}$, $i=1,\cdots ,m$. In light of Lemma 2
and by the induction, we obtain
\begin{equation}
|\pi \psi \rangle _{1\cdots n}=|\psi _{1}^{\prime }\rangle _{R_{1}}|\psi
_{2}^{\prime }\rangle _{R_{2}}\cdots |\psi _{m}^{\prime }\rangle _{R_{m}}.
\end{equation}%
Then, we obtain
\begin{eqnarray}
&&|\psi \rangle _{1\cdots n}  \notag \\
&=&|\pi ^{-1}\psi _{1}^{\prime }\rangle _{\pi ^{-1}(R_{1})}|\pi ^{-1}\psi
_{2}^{\prime }\rangle _{\pi ^{-1}(R_{2})}\cdots |\pi ^{-1}\psi _{m}^{\prime
}\rangle _{\pi ^{-1}(R_{m})}.  \notag \\
&&
\end{eqnarray}%
Let $\pi ^{-1}(R_{i})=S_{i}$ and we write $|\pi ^{-1}\psi _{i}^{\prime
}\rangle $ as $|\psi _{i}\rangle $, $i=1,\cdots ,m$. Then, Eq. (\ref{g-n-7})
holds.

Theorem 3 and Lemma 5 lead to the following theorem.

\textit{Theorem 4.}
\begin{equation}
|\psi \rangle _{1\cdots n}=|\psi _{1}\rangle _{S_{1}}|\psi _{2}\rangle
_{S_{2}}\cdots |\psi _{m}\rangle _{S_{m}},
\end{equation}%
where $n$ qubits are split into $m$ parts $S_{1}$, $S_{2}$, $\cdots $, $%
S_{m} $, if and only if there exists a permutation of qubits, for example $%
\pi $, such that the coefficient vector $C(|\pi \psi \rangle )$ can be
decomposed into

\begin{equation}
C(|\pi \psi \rangle )=V_{1}\otimes V_{2}\otimes \cdots \otimes V_{m}.
\end{equation}

\section{The coefficient vector decomposition and the proportionality of two
block vectors}

In this section, we will use the proportionality of two vectors to solve the
tensor equation in Eq. (\ref{tensor-eq-1}). We define that a vector $v$ is
proportional to a non-zero vector $u$ if $v=ku$, where $k$ is a complex
number. Specially, when $v=0$, then $k=0$. Thus, a zero-vector is always
proportional to a non-zero vector.

To solve Eq. (\ref{tensor-eq-1}), we divide the coefficient vector $C(|\psi
\rangle )$ of the state $|\psi \rangle _{1\cdots n}$\ of $n$ qubits into $%
2^{i}$ equal-size block vectors $C^{(\ell )}$, $\ell =0,1,\cdots ,2^{i}-1$.
Then $C(|\psi \rangle )$\ can be written as
\begin{equation}
C(|\psi \rangle )=\left(
\begin{array}{c}
C^{(0)} \\
C^{(1)} \\
\vdots \\
C^{(2^{i}-1)}%
\end{array}%
\right) ,  \label{bl-1}
\end{equation}%
where $C^{(\ell )}=\left(
\begin{array}{cccc}
c_{\ell \times 2^{n-i}} & c_{\ell \times 2^{n-i}+1} & \cdots & c_{\ell
\times 2^{n-i}+2^{n-i}-1}%
\end{array}%
\right) ^{T}$, $\ell =0,1,\cdots ,2^{i}-1$. One can know that each block
vector $C^{(\ell )}$ is a column vector of the size $2^{n-i}$. Since $%
C(|\psi \rangle )\neq 0$, so at least one of the block vectors $C^{(\ell )}$%
, $\ell =0,1,\cdots ,2^{i}-1$, does not vanish.

Next, we will show that to detect if a state is genuinely entangled,
factually\ we don't need to solve Eq. (\ref{tensor-eq-1}), we only need to
know if Eq. (\ref{tensor-eq-1}) has a solution.\

\subsection{A necessary and sufficient condition for the coefficient vector
decomposition via the proportionality of two block vectors}

\textit{Lemma 6.} If the coefficient vector $C(|\psi \rangle )$\ of a pure
state $|\psi \rangle _{1\cdots n}$\ of $n$ qubits can be decomposed into

\begin{equation}
C(|\psi \rangle )=V_{1}\otimes V_{2},  \label{coef-ten-1}
\end{equation}%
where $V_{1}$ and $V_{2}$ are $2^{i}$ and $2^{n-i}$\ dimensional vectors,
respectively, then any two non-zero block vectors of the block matrix in Eq.
(\ref{bl-1})\ are proportional.

Proof. Let $V_{1}=(a_{0},a_{1},\cdots ,a_{2^{i}-1})^{T}$. Then, from Eqs. (%
\ref{bl-1}, \ref{coef-ten-1}) we obtain
\begin{equation}
C^{(\ell )}=a_{\ell }V_{2},  \label{eq-1-}
\end{equation}%
where $\ell =0,\cdots ,2^{i}-1$. Let $C^{(k)}$ be any non-zero block vector.
Then,$\ a_{k}\neq 0$ and $V_{2}=\frac{1}{a_{k}}C^{(k)}$. Then, from Eq. (\ref%
{eq-1-}) we obtain
\begin{equation}
C^{(\ell )}=\frac{a_{\ell }}{a_{k}}C^{(k)},  \label{eq-3}
\end{equation}%
where $\ell =0,1,\cdots ,2^{i}-1$. Clearly, all the block vectors $C^{(\ell
)}$ are proportional to the non-zero block vector $C^{(k)}$. Note that $%
C^{(k)}$ is any non-zero block vector. Therefore, any two non-zero block
vectors of the block matrix in Eq. (\ref{bl-1})\ are proportional.

\textit{Lemma} 7. For the block matrix in Eq. (\ref{bl-1}), if there is a
non-zero block vector such that all the block vectors are proportional to
the non-zero block vector, then $C(|\psi \rangle )=V_{1}\otimes V_{2}$,
where $V_{1}$ and $V_{2}$ are $2^{i}$ and $2^{n-i}$\ dimensional vectors,
respectively.

Proof. Without loss of generality, assume that the block vector $C^{(0)}\neq
0$ and all the block vectors $C^{(\ell )}$ are proportional to $C^{(0)}$.
Thus, we can write
\begin{equation}
C^{(\ell )}=k_{\ell }C^{(0)},\ell =0,1,\cdots ,2^{i}-1.
\end{equation}%
Then, it is easy to see that
\begin{equation}
C(|\psi \rangle )=\left(
\begin{array}{c}
C^{(0)} \\
C^{(1)} \\
\vdots \\
C^{(2^{i}-1)}%
\end{array}%
\right) =\left(
\begin{array}{c}
1 \\
k_{1} \\
\vdots \\
k_{2^{i}-1}%
\end{array}%
\right) \otimes C^{(0)}.
\end{equation}%
That is, $C(|\psi \rangle )$ is a tensor product of two column vectors.

Lemmas 6 and 7 lead to the following theorems.

\textit{Theorem 5.} The coefficient vector $C(|\psi \rangle )$\ of a pure
state $|\psi \rangle _{1\cdots n}$\ of $n$ qubits can be decomposed into

\begin{equation}
C(|\psi \rangle )=V_{1}\otimes V_{2},  \label{ten-dec-1}
\end{equation}%
where $V_{1}$ and $V_{2}$ are $2^{i}$ and $2^{n-i}$\ dimensional vectors,
respectively, if and only if any two non-zero block vectors of the block
matrix in Eq. (\ref{bl-1})\ are proportional.

\textit{Theorem 6.} The coefficient vector $C(|\psi \rangle )$\ of a pure
state $|\psi \rangle _{1\cdots n}$\ of $n$ qubits can be decomposed into

\begin{equation}
C(|\psi \rangle )=V_{1}\otimes V_{2},
\end{equation}%
where $V_{1}$ and $V_{2}$ are $2^{i}$ and $2^{n-i}$\ dimensional vectors,
respectively, if and only if for the block matrix in Eq. (\ref{bl-1}), there
is a non-zero block vector such that all the block vectors are proportional
to the non-zero block vector.

In light of Theorem 5, we obtain the following theorem for the genuine
entanglement.

\textit{Theorem 7.} (For genuine entanglement) A pure state $|\psi \rangle
_{1\cdots n}$\ of $n$ qubits is genuinely entangled if and only if for any
permutation $\pi $\ of qubits in Appendix A and the block matrix of $C(\pi
|\psi \rangle )$ in Eq. (\ref{bl-1}), there are two non-zero block vectors
which are not proportional.

In light of Theorem 5, we obtain the following corollaries.

\textit{Corollary 2}. A pure product state $|\psi \rangle _{1\cdots n}$\ of $%
n$ qubits is of the form $|\psi \rangle _{1\cdots n}=|\varphi \rangle
_{1}|\phi \rangle _{2\cdots n}$ if and only if for the block matrix
\begin{equation}
C(|\psi \rangle )=\left(
\begin{array}{c}
C^{(0)} \\
C^{(1)}%
\end{array}%
\right) ,
\end{equation}%
the block vectors $C^{(0)}$ and $C^{(1)}$ are proportional. Note that $%
C^{(0)}=(c_{0},c_{1},\cdots ,c_{2^{n-1}-1})^{T}$ and $%
C^{(1)}=(c_{2^{n-1}},c_{2^{n-1}+1},\cdots ,c_{2^{n}-1})^{T}$.

To detect if $|\psi \rangle _{1\cdots n}=|\varphi \rangle _{i}|\phi \rangle
_{1\cdots i(i+1)\cdots n}$, we only need the following tests.

\textit{Corollary 3.} Let $\pi $ be a transposition $(i,1)$\ of qubits 1 and
$i$, where $i=1,\cdots ,n$. $|\psi \rangle _{1\cdots n}=|\varphi \rangle
_{i}|\phi \rangle _{1\cdots i(i+1)\cdots n}$ if and only if the two column
vectors $(c_{0},c_{1},\cdots ,c_{2^{n-1}-1})^{T}$ and $%
(c_{2^{n-1}},c_{2^{n-1}+1},\cdots ,c_{2^{n}-1})^{T}$ are proportional, where
$c_{k}$ are the coefficients of the state $|\pi \psi \rangle $.

Corollary 4. Let $|\psi \rangle _{123}=\sum c_{i}|i\rangle $ be any pure
state of three qubits. Then, $|\psi \rangle _{123}$ is product if and only
if $\left(
\begin{array}{c}
c_{0} \\
c_{1} \\
c_{2} \\
c_{3}%
\end{array}%
\right) \ $and $\left(
\begin{array}{c}
c_{4} \\
c_{5} \\
c_{6} \\
c_{7}%
\end{array}%
\right) $, $\left(
\begin{array}{c}
c_{0} \\
c_{1} \\
c_{4} \\
c_{5}%
\end{array}%
\right) \ $and $\left(
\begin{array}{c}
c_{2} \\
c_{3} \\
c_{6} \\
c_{7}%
\end{array}%
\right) $, or $\left(
\begin{array}{c}
c_{0} \\
c_{2} \\
c_{4} \\
c_{6}%
\end{array}%
\right) \ $and $\left(
\begin{array}{c}
c_{1} \\
c_{3} \\
c_{5} \\
c_{7}%
\end{array}%
\right) $ are proportional. Otherwise, it is genuinely entangled.

For example, for GHZ (resp. W, $|\zeta \rangle =\frac{1}{2}(|001\rangle
+|010\rangle +|100\rangle +|111\rangle )$), all the above three pairs of
block vectors become $\frac{1}{\sqrt{2}}\left(
\begin{array}{c}
1 \\
0 \\
0 \\
0%
\end{array}%
\right) $ and $\frac{1}{\sqrt{2}}\left(
\begin{array}{c}
0 \\
0 \\
0 \\
1%
\end{array}%
\right) $ (resp. $\frac{1}{\sqrt{3}}\left(
\begin{array}{c}
0 \\
1 \\
1 \\
0%
\end{array}%
\right) $ and $\frac{1}{\sqrt{3}}\left(
\begin{array}{c}
1 \\
0 \\
0 \\
0%
\end{array}%
\right) $, $\frac{1}{2}\left(
\begin{array}{c}
0 \\
1 \\
1 \\
0%
\end{array}%
\right) $ and $\frac{1}{2}\left(
\begin{array}{c}
1 \\
0 \\
0 \\
1%
\end{array}%
\right) $). In light of Corollary 4, GHZ, W and $|\zeta \rangle $ are
genuinely entangled.

\subsection{Solving the tensor equation in Eq. (\protect\ref{tensor-eq-1})
via the proportionality of two block vectors}

Let $V_{1}=(a_{0},a_{1},\cdots ,a_{2^{i}-1})^{T}$. Then, Eq. (\ref%
{tensor-eq-1}) reduces to
\begin{equation}
C^{(\ell )}=a_{\ell }V_{2},  \label{ten-e-1}
\end{equation}%
where $\ell =0,1,\cdots ,2^{i}-1$. Specially,
\begin{equation}
C^{(0)}=a_{0}V_{2}.  \label{ten-e-2}
\end{equation}

Without loss of generality, assume that the block vector $C^{(0)}\neq 0$. We
use $||v||$ to stand for $L^{2}$ norm of a vector $v$. Here, let $%
||V_{2}||=1 $, which means that the state $|\phi \rangle _{(i+1)\cdots n}$
in Eq. (\ref{separ-1})\ is a normalized pure state. Then, from Eq. (\ref%
{ten-e-2}) we obtain $|a_{0}|=||C^{(0)}||$. Without loss of generality,
ignoring the phase factor let
\begin{equation}
a_{0}=||C^{(0)}||.  \label{a0-v}
\end{equation}%
Then, via Eq. (\ref{ten-e-2}) we obtain
\begin{equation}
V_{2}=\frac{1}{a_{0}}C^{(0)}=\frac{C^{(0)}}{||C^{(0)}||}.  \label{ten-e-3}
\end{equation}

Next we solve $V_{1}$. Via Eq. (\ref{ten-e-3}), Eq. (\ref{ten-e-1})\ becomes
\begin{equation}
C^{(\ell )}=\frac{a_{\ell }}{a_{0}}C^{(0)},  \label{ten-e-4}
\end{equation}%
where $\ell =0,1,\cdots ,2^{i}-1$. Note that the block vectors $C^{(\ell )}$%
, $\ell =0,1,\cdots ,2^{i}-1$, are given. From Eq. (\ref{ten-e-4}), one can
know that $a_{\ell }$ has a solution if and only if the block vector $%
C^{(\ell )}$ is proportional to the block vector $C^{(0)}$. Assume that $%
C^{(\ell )}=k_{\ell }C^{(0)}$, $\ell =1,\cdots ,2^{i}-1$. Then, $a_{\ell
}=k_{\ell }a_{0}$ and we obtain
\begin{equation}
V_{1}=\left(
\begin{array}{c}
a_{0} \\
a_{1} \\
\vdots \\
a_{2^{i}-1}%
\end{array}%
\right) =\left(
\begin{array}{c}
a_{0} \\
k_{1}a_{0} \\
\vdots \\
k_{2^{i}-1}a_{0}%
\end{array}%
\right) .  \label{ten-e-5}
\end{equation}%
Thus, Eq. (\ref{tensor-eq-1}) has a solution in Eqs (\ref{a0-v}, \ref%
{ten-e-3}, \ref{ten-e-5}). That is, the coefficient vector can be
decomposed. If some block vector $C^{(\ell )}$ is not proportional to the
block vector $C^{(0)}$, then in light of Theorem 5, Eq. (\ref{tensor-eq-1})
does not have a solution. That is, the state $|\psi \rangle _{1\cdots n}$
cannot be written as$\ |\psi \rangle _{1\cdots n}=|\varphi \rangle _{1\cdots
i}|\phi \rangle _{(i+1)\cdots n}$.

\subsection{Complexity}

It is known that a product state of four qubits is of the following forms: $%
|\varphi \rangle _{1}|\phi \rangle _{234}$, $|\varphi \rangle _{2}|\phi
\rangle _{134}$, $|\varphi \rangle _{3}|\phi \rangle _{124}$, $|\varphi
\rangle _{4}|\phi \rangle _{123}$, $|\varphi \rangle _{12}|\phi \rangle
_{34} $, $|\varphi \rangle _{13}|\phi \rangle _{24}$, $|\varphi \rangle
_{14}|\phi \rangle _{23}$. To decide if a state of four qubits is genuinely
entangled, we need to show that the state cannot be written as any one of
the above forms. In light of Theorem 1, we can decide if a state of $n$
qubits can be written as $|\varphi \rangle _{1\cdots i}|\phi \rangle
_{(i+1)\cdots n}$ by decomposing the coefficient vector.

For four qubits, one can see that by applying the transposition $(1,2)$ of
qubits 1 and 2 (resp. $(1,3)$ of qubits 1 and 3, $(1,4)$ of qubits 1 and 4),
the state $|\varphi \rangle _{2}|\phi \rangle _{134}$ (resp. $|\varphi
\rangle _{3}|\phi \rangle _{124}$, $|\varphi \rangle _{4}|\phi \rangle
_{123} $) becomes $|\varphi \rangle _{1}|\phi \rangle _{234}$, and by
applying the transpositions $(2,3)$ of qubits 2 and 3 and $(2,4)$ of qubtis
2 and 4, the states $|\varphi \rangle _{13}|\phi \rangle _{24}$ and $%
|\varphi \rangle _{14}|\phi \rangle _{23}$ become $|\varphi \rangle
_{12}|\phi \rangle _{34}$. Applying permutations of qubits in Appendix A,
any pure product state of $n $ qubits becomes $|\varphi \rangle _{1\cdots
i}|\phi \rangle _{(i+1)\cdots n} $. For example, let $\pi $ be the
transposition $(1,2)$ of qubits 1 and 2. Then, if the coefficient vector of $%
\pi |\psi \rangle _{1\cdots n}$ can be decomposed into $\pi |\psi \rangle
_{1\cdots n}=V_{1}\otimes V_{2}$, where $V_{1}$ and $V_{2}$ are $2$ and $%
2^{n-1}$\ dimensional vectors, respectively, then $\pi |\psi \rangle
_{1\cdots n}$ is a product state of the form $|\varphi ^{\prime }\rangle
_{1}|\phi ^{\prime }\rangle _{23\cdots n}$ in light of Theorem 1. Then, $%
|\psi \rangle _{1\cdots n}$ is a product state of the form $|\varphi \rangle
_{2}|\phi \rangle _{13\cdots n}$.

Therefore, to detect if a state of $n$ qubits is genuinely entangled, in
worst case we need $2^{n-1}-(n-1)$ permutations of qubits in Appendix A and
we need to check if there are two non-zero block vectors which are not
proportional for $2^{n-1}-1$ block matrices in Eq. (\ref{bl-1}).

\section{Comparison}

In our paper, we study the separability of pure states of $n$ qubits. We
give a necessary and sufficient condition\ for the separability of pure
states of $n$ qubits via the coefficient vector decomposition and
permutations of qubits. We compare our work to the previous works below and
in Table II.

In II. B of \cite{Verstraete01}, the authors constructed an $N_{1}\times
N_{2}$ matrix whose entries are the rearranging coefficients of a pure state
of an $N_{1}\times N_{2}$ bipartite system. Then, they showed that a pure
state of an $N_{1}\times N_{2}$ bipartite system is separable if and only if
all $2\times 2$ minors of the $N_{1}\times N_{2}$ matrix must be zero.

The above necessary and sufficient condition holds for only bipartite
systems. It is easy to check that the condition is not suitable for
multipartite systems including $n$-qubit systems. For example, for the
product state of four qubits:$\frac{1}{2}(|00\rangle +|11\rangle )\otimes
(|00\rangle +|11\rangle )$, some $2\times 2$ minors of the coefficient
matrix do not vanish.

In Fact 1 of \cite{Vicente}, via correlation tensors the authors
demonstrated that a pure state of a multipartite system\ is biseparable if
and only if all the $m$-body correlation tensors factorize into the
corresponding $k$-body correlation tensor of the $k$ particles and the $%
(m-k) $-body correlation tensor of the $m-k$ particles.

Clearly, decomposing a vector into a tensor product of two vectors is easier
than decomposing a correlation tensor.

In \cite{Yu}, the authors gave a necessary and sufficient condition\ for the
full separability of pure states of three qubits in Lemma 1 and Theorem 1.
Their Theorem 1 stated that a pure state $|\psi \rangle $ of three qubits is
fully separable iff $\langle \psi ^{\ast }|s^{\alpha }|\psi \rangle =0$, $%
\alpha =1,\cdots ,9$, ref. \cite{Yu} for the definitions of $s^{\alpha }$.

Comparably, for a pure state $|\psi \rangle $\ of three qubits, our approach
only needs to decompose two coefficient vectors $C(|\psi \rangle )$ and $%
C(\pi |\psi \rangle )$, where $\pi $ is the transposition (1,2) of qubits 1
and 2.

Theorem 1 of \cite{Bastin18} claimed that a general N-partite pure state $%
|\psi \rangle $ is separable if and only if the generalized concurrences $%
C_{\alpha }(\psi )=0$, $\alpha =1,\cdots ,Q$.

For an $n$-qubit system, $Q=nd^{n}(d^{n-1}-1)(d-1)/4$ \cite{Bastin18}. For
qubits 2 to 10, they calculated the number $Q$ of the concurrences (see
Table I). While we need to check if there are two non-zero block vectors
which are not proportional for $D$ ($=2^{n-1}-1$) block matrices in Eq. (\ref%
{bl-1}). Clearly, $D\ll Q$.\ For qubits 2 to 10, we list the values of $D$
in Table I.

Comparably, (1) the necessary and sufficient condition in this paper is
simpler and more intuitional than the previous conditions;\ (2) via the
decomposition of a coefficient vector of a pure state, in light of Theorem 4
we can straightforwardly and constructively factorize the state. Comparably,
it is not convenient to factorize a state via the conditions\cite{Vicente,
Verstraete01, Bastin18, Yu}; (3) we derive the full decomposition theorem.

Table I. compare for qubits 2 to 10

$%
\begin{tabular}{|c|c|c|c|c|c|c|}
\hline
qubits & 2 & 3 & 4 & 5 & 6 & 7 \\ \hline
$Q$ & 2 & 18 & 112 & 600 & 2976 & 14112 \\ \hline
$D$ & 1 & 3 & 7 & 15 & 31 & 63 \\ \hline
qubits & 8 & 9 & 10 &  &  &  \\ \hline
$Q$ & 65024 & 293760 & 1308160 &  &  &  \\ \hline
$D$ & 127 & 255 & 511 &  &  &  \\ \hline
\end{tabular}%
$

Table II. The conditions for the separability of pure states

\begin{tabular}{|c|c|c|}
\hline
ref. & system & a pure state is separable iff \\ \hline
\cite{Verstraete01} & bipartite & all minors of the matrix vanish \\ \hline
\cite{Yu} & 3-qubits & $\langle \psi ^{\ast }|s^{\alpha }|\psi \rangle =0$, $%
\alpha =1,\cdots ,9$ \\ \hline
\cite{Vicente} & multiparti & all the correlation tensors factorize \\ \hline
\cite{Bastin18} & multiparti & generalized concurrences $C_{\alpha }(\psi )=0
$ \\ \hline
this & n qubits & the coefficient vector factorizes \\ \hline
\end{tabular}

\section{Summary}

In this paper, we propose a necessary and sufficient condition for detecting
the genuine entanglement for $n$ qubits\ via the tensor product
decomposition of the coefficient vectors and permutations of qubits.
Moreover, we show that a coefficient vector can be decomposed if and only if
any two non-zero block vectors of the block matrix of the coefficient vector
are proportional. Thus, we avoid decomposing coefficient vectors to detect
the genuine entanglement.

We also give the full decomposition theorem. We split $n$ qubits into $m$
parts $S_{1}$, $S_{2}$, $\cdots $, $S_{m}$. If a pure state of $n$ qubits is
a product of states of qubits in $S_{i}$, $i=1,\cdots ,m$, then there exists
a permutation of qubits, for example $\pi $, such that the coefficient
vector $C(|\pi \psi \rangle )$ can be decomposed into a tensor product of $m$
vectors.

For symmetric states of $n$ qubits \cite{Bastin}, the coefficient vectors do
not change under any permutations of qubits. So, for detecting the genuine
entanglement of symmetric states we only need to check the proportionality
of block vectors of the block matrices of the coefficient vectors without
considering permutations of qubits. For example, it is easy to check that
GHZ, W, Dicke states $|i,n\rangle $, GHZ+W, GHZ+Dicke states of $n$ qubits
are genuinely entangled.

Acknowledgement---This work was supported by Tsinghua National Laboratory
for Information Science and Technology.

\section{Appendix A. The number of the vectors decomposed}

\subsection{Odd $n$ qubits}

\subsubsection{Three qubits}

For three qubits, there are three kinds of pure product states. Case 1. $%
|\psi \rangle _{123}=|\varphi \rangle _{1}|\phi \rangle _{23}$; Case 2. $%
|\psi \rangle _{123}=|\varphi \rangle _{2}|\phi \rangle _{13}$; Case 3. $%
|\psi \rangle _{123}=|\varphi \rangle _{3}|\phi \rangle _{12}$. For Cases 1
and 3, in light of Lemma 1 their coefficient vectors $C(|\psi \rangle )$\
can be decomposed into a tensor product of two vectors. For Case 2, we need
to apply the transposition $\pi =(1,2)$ of qubits 1 and 2 to $|\psi \rangle
_{123}$. Then $|\psi \rangle _{123}$ become $|\pi \psi \rangle
_{123}=|\varphi ^{\prime }\rangle _{1}|\phi ^{\prime }\rangle _{23}$ and the
vector $C(|\pi \psi \rangle )$ can be decomposed. For Case 2, we need $%
\left(
\begin{array}{c}
3 \\
1%
\end{array}%
\right) -2$ $(=1)$ permutation. In total, we need to decompose $1+\left(
\begin{array}{c}
3 \\
1%
\end{array}%
\right) -2$ $(=2)$ vectors.

\subsubsection{Five qubits}

For five qubits, there are the following pure product states. Cases:

1. $|\psi \rangle _{12345}=|\varphi \rangle _{1}|\phi \rangle _{2345}$; 2. $%
|\psi \rangle _{12345}=|\varphi \rangle _{2}|\phi \rangle _{1345}$; 3. $%
|\psi \rangle _{12345}=|\varphi \rangle _{3}|\phi \rangle _{1245}$; 4. $%
|\psi \rangle _{12345}=|\varphi \rangle _{4}|\phi \rangle _{1235}$; 5. $%
|\psi \rangle _{12345}=|\varphi \rangle _{5}|\phi \rangle _{1234}$;

6. $|\psi \rangle _{12345}=|\varphi \rangle _{12}|\phi \rangle _{345}$; 7. $%
|\psi \rangle _{12345}=|\varphi \rangle _{13}|\phi \rangle _{245}$; 8. $%
|\psi \rangle _{12345}=|\varphi \rangle _{14}|\phi \rangle _{235}$; 9. $%
|\psi \rangle _{12345}=|\varphi \rangle _{15}|\phi \rangle _{234}$; 10. $%
|\psi \rangle _{12345}=|\varphi \rangle _{23}|\phi \rangle _{145}$; 11. $%
|\psi \rangle _{12345}=|\varphi \rangle _{24}|\phi \rangle _{135}$; 12. $%
|\psi \rangle _{12345}=|\varphi \rangle _{25}|\phi \rangle _{134}$; 13. $%
|\psi \rangle _{12345}=|\varphi \rangle _{34}|\phi \rangle _{125}$; 14. $%
|\psi \rangle _{12345}=|\varphi \rangle _{35}|\phi \rangle _{124}$; 15. $%
|\psi \rangle _{12345}=|\varphi \rangle _{45}|\phi \rangle _{123}$.

For Cases 1, 5, 6, and 15, in light of Lemma 1 their coefficient vectors can
be decomposed. For Cases 2-4, we need $\left(
\begin{array}{c}
5 \\
1%
\end{array}%
\right) -2$ permutations to decompose their coefficient vectors. For Cases
7-14, we need $\left(
\begin{array}{c}
5 \\
2%
\end{array}%
\right) -2$ permutations. In total, we need to decompose $1+\left(
\begin{array}{c}
5 \\
1%
\end{array}%
\right) -2+\left(
\begin{array}{c}
5 \\
2%
\end{array}%
\right) -2$ ($=12$) vectors.

\subsubsection{Odd $n$ qubits}

For any odd $n$ qubits, we calculate the number of the vectors decomposed
below.

Case 1. $|\psi \rangle =|\varphi \rangle _{i_{1}}|\phi \rangle _{i_{2}\cdots
i_{n}}$. For $|\varphi \rangle _{1}|\phi \rangle _{2\cdots n}$ and $|\varphi
\rangle _{n}|\phi \rangle _{1\cdots (n-1)}$, in light of Lemma 1 their
coefficient vectors can be decomposed. For $i_{1}=2,\cdots ,(n-1)$, let the
transposition $\pi =(1,i_{1})$ of qubits 1 and $i_{1}$. Then, $\pi |\psi
\rangle =|\varphi ^{\prime }\rangle _{1}|\phi ^{\prime }\rangle _{2\cdots n}$%
. In light of Lemma 1, the coefficient vectors of $\pi |\psi \rangle $ can
be decomposed. Thus, we need $\left(
\begin{array}{c}
n \\
1%
\end{array}%
\right) -2$ permutations.

Case 2. $|\psi \rangle =|\varphi \rangle _{i_{1}i_{2}}|\phi \rangle
_{i_{3}\cdots i_{n}}$. Assume that $i_{1}<i_{2}$. For $|\varphi \rangle
_{12}|\phi \rangle _{3\cdots n}$ and $|\varphi \rangle _{1\cdots (n-2)}|\phi
\rangle _{(n-1)n}$, in light of Lemma 1 their coefficient vectors can be
decomposed. Then, let the permutation $\sigma =(i_{2},2)(i_{1},1)$. Then, $%
\sigma |\psi \rangle =|\varphi ^{\prime }\rangle _{12}|\phi ^{\prime
}\rangle _{3\cdots n}$. So, we need $\left(
\begin{array}{c}
n \\
2%
\end{array}%
\right) -2$ different permutations.

Generally,%
\begin{eqnarray*}
&&1+\left(
\begin{array}{c}
n \\
1%
\end{array}%
\right) -2+\cdots +\left(
\begin{array}{c}
n \\
(n-1)/2%
\end{array}%
\right) -2 \\
&=&1+\frac{1}{2}\left( \sum_{i=0}^{n}\left(
\begin{array}{c}
n \\
i%
\end{array}%
\right) -2\right) -(n-1) \\
&=&2^{n-1}-(n-1).
\end{eqnarray*}

In total, we need to decompose $2^{n-1}-(n-1)$ vectors for odd $n$ qubits.

\subsection{Even $n$ qubits}

\subsubsection{Four qubits}

For four qubits, we have the following pure product states. Cases: 1. $|\psi
\rangle _{1234}=|\varphi \rangle _{1}|\phi \rangle _{234}$; 2. $|\psi
\rangle _{1234}=|\varphi \rangle _{2}|\phi \rangle _{134}$; 3. $|\psi
\rangle _{1234}=|\varphi \rangle _{3}|\phi \rangle _{124}$; 4. $|\psi
\rangle _{1234}=|\varphi \rangle _{4}|\phi \rangle _{123}$; 5. $|\psi
\rangle _{1234}=|\varphi \rangle _{12}|\phi \rangle _{34}$; 6. $|\psi
\rangle _{1234}=|\varphi \rangle _{13}|\phi \rangle _{24}$; 7. $|\psi
\rangle _{1234}=|\varphi \rangle _{14}|\phi \rangle _{23}$.

For Cases 1, 4, and 5, their coefficient vectors $C(|\psi \rangle )$\ can be
decomposed into a tensor product of two vectors. We need $\left(
\begin{array}{c}
4 \\
1%
\end{array}%
\right) -2$ permutations for Cases 2 and 3 to decompose their coefficient
vectors. We need $\frac{1}{2}\left(
\begin{array}{c}
4 \\
2%
\end{array}%
\right) -1$ permutations for Cases 6 and 7. In total, we need to decompose $%
1+\left(
\begin{array}{c}
4 \\
1%
\end{array}%
\right) -2+\frac{1}{2}\left(
\begin{array}{c}
4 \\
2%
\end{array}%
\right) -1$ ($=5$) vectors.

\subsubsection{Even $n$ qubits}

For any even $n$ qubits, we calculate the number of the vectors decomposed
below.

Case 1. $|\psi \rangle =|\varphi \rangle _{i_{1}}|\phi \rangle _{i_{2}\cdots
i_{n}}$, where $i_{1}=1,\cdots ,n$. Similarly, we need $\left(
\begin{array}{c}
n \\
1%
\end{array}%
\right) -2$ permutations. Case 2. $|\psi \rangle =|\varphi \rangle
_{i_{1}i_{2}}|\phi \rangle _{i_{3}\cdots i_{n}}$. Similarly, we need $\left(
\begin{array}{c}
n \\
2%
\end{array}%
\right) -2$ permutations. Case 3. $|\psi \rangle =|\varphi \rangle
_{i_{1}i_{2}\cdots i_{(n/2)}}|\phi \rangle _{i_{(n/2)+1}\cdots i_{n}}$. We
need $\frac{1}{2}\left(
\begin{array}{c}
n \\
n/2%
\end{array}%
\right) -1$ permutations. Then, a calculation yields the number of the
vectors decomposed below.
\begin{eqnarray*}
&&1+\left(
\begin{array}{c}
n \\
1%
\end{array}%
\right) -2+\cdots \left(
\begin{array}{c}
n \\
n/2-1%
\end{array}%
\right) -2+\frac{1}{2}\left(
\begin{array}{c}
n \\
n/2%
\end{array}%
\right) -1 \\
&=&\frac{1}{2}\left[ 2\left(
\begin{array}{c}
n \\
1%
\end{array}%
\right) +\cdots +2\left(
\begin{array}{c}
n \\
n/2-1%
\end{array}%
\right) +\left(
\begin{array}{c}
n \\
n/2%
\end{array}%
\right) \right] +2-n \\
&=&\frac{1}{2}(2^{n}-2)+2-n \\
&=&2^{n-1}-(n-1)
\end{eqnarray*}

\end{document}